\documentclass[12pt,preprint]{aastex6}
\usepackage{amsmath}
\usepackage{graphicx}
\usepackage{enumitem}
\usepackage{color}

\shorttitle{Decay-less oscillations and seismology of transverse MHD waves}
\shortauthors{P. Antolin et al.}

\begin{document}

\title{Modelling observed decay-less oscillations as resonantly enhanced Kelvin-Helmholtz vortices from transverse MHD waves and their seismological application}

\author{P. Antolin\altaffilmark{1,4}, I. De Moortel\altaffilmark{1}, T. Van Doorsselaere\altaffilmark{2}, T. Yokoyama\altaffilmark{3}}
\affil{\altaffilmark{1}School of Mathematics and Statistics, University of St. Andrews, St. Andrews, Fife KY16 9SS, UK\\
\altaffilmark{2}Centre for mathematical Plasma Astrophysics, Mathematics Department, KU Leuven, Celestijnenlaan 200B bus 2400, B-3001 Leuven, Belgium\\
\altaffilmark{3}The University of Tokyo, Hongo, Bunkyo-ku, Tokyo 113-0033, Japan\\
\altaffilmark{4}National Astronomical Observatory of Japan, Osawa, Mitaka, Tokyo 181-8588, Japan}
\email{patrick.antolin@st-andrews.ac.uk}

\begin{abstract}

In the highly structured solar corona, resonant absorption is an unavoidable mechanism of energy transfer from global transverse MHD waves to local azimuthal Alfv\'en waves. Due to its localised nature, a direct detection of this mechanism is extremely difficult. Yet, it is the leading theory explaining the observed fast damping of the global transverse waves. However, at odds with this theoretical prediction, recent observations indicate that in the low amplitude regime such transverse MHD waves can also appear decay-less, a yet unsolved phenomenon. Recent numerical work has shown that Kelvin-Helmholtz instabilities (KHI) often accompany transverse MHD waves. In this work, we combine 3D MHD simulations and forward modelling to show that for currently achieved spatial resolution and observed small amplitudes, an apparent decay-less oscillation is obtained. This effect results from the combination of periodic brightenings produced by the KHI and the coherent motion of the KHI vortices amplified by resonant absorption. Such effect is especially clear in emission lines forming at temperatures that capture the boundary dynamics rather than the core, and reflects the low damping character of the local azimuthal Alfv\'en waves resonantly coupled to the kink mode. Due to phase mixing, the detected period can vary depending on the emission line, with those sensitive to the boundary having shorter periods than those sensitive to the loop core. This allows to estimate the density contrast at the boundary. 

\end{abstract}


\section{Introduction}

Research in the last decade has shown that waves and oscillations permeate the solar atmosphere and constitute coronal heating candidates. Of particular interest among these waves are transverse MHD waves. Their characteristic fast damping, particularly for strong amplitudes \citep{Aschwanden_1999ApJ...520..880A,Nakariakov_1999Sci...285..862N,Arregui_2012LRSP....9....2A, DeMoortel_Nakariakov_2012RSPTA.370.3193D,Verwichte_2013AA...552A.138V,Goddard_2016AA...590L...5G} is successfully explained by resonant absorption and mode coupling. 

Resonant absorption (or mode coupling for propagating waves) is an ideal process of energy transfer between different wave modes \citep{Ionson_1978ApJ...226..650I, Goossens_2002AA...394L..39G,Goossens_2011SSRv..158..289G,Pascoe_2010ApJ...711..990P,DeMoortel_2016PPCF...58a4001D}, which has been shown to be very efficient and robust \citep{DeMoortel_Nakariakov_2012RSPTA.370.3193D,Pascoe_etal_2011ApJ...731...73P,Terradas_2008ApJ...679.1611T}. This mechanism predicts that in the classical picture of coronal loops with a smooth density gradient between the inside and the outside of the flux tube the global transverse mode can resonantly couple to local Alfv\'en waves of azimuthal character. The global transverse mode, which consists of a purely transverse displacement of the loop core, then behaves as azimuthal Alfv\'en waves at the boundary for most of the oscillation time. This means that most of the displacement (and thus the energy) in such waves is azimuthal and local rather than transverse and global, making resonant absorption extremely difficult to observe directly. 

Recently, decay-less transverse oscillations have been reported \citep{Nistico_2013AA...552A..57N, Anfinogentov_2013AA...560A.107A, Anfinogentov_2015AA...583A.136A,Goddard_2016AA...585A.137G}, apparently at odds with resonant absorption theory. These events have a rather ubiquitous character in active regions and correspond well to fundamental (standing) kink modes. The clearly decay-less cases seem to occur on long loops with small perturbation amplitudes that on average are less than $1\%$ of the kink speed.  
 
Numerical simulations in 3D MHD have shown that transverse MHD waves can become unstable to KHI due to shear motions at the boundary of flux tubes \citep{Karpen_1993ApJ...403..769K,Ofman_1994GeoRL..21.2259O,Poedts_1997SoPh..172...45P, Terradas_2008ApJ...687L.115T,Antolin_2014ApJ...787L..22A, Zaqarashvili_2015ApJ...813..123Z}. The KHI associated with transverse MHD waves leads to the generation of a myriad of vortices and current sheets along the flux tube, so called TWIKH (Transverse Wave Induced Kelvin-Helmholtz) rolls. The mixing and the heating produced by the KHI, combined with the compressive nature of the vortices, leads to strand-like structure in intensity images in coronal loops \citep{Antolin_2014ApJ...787L..22A}, and thread-like structure in prominences \citep{Antolin_2015ApJ...809...72A}. Furthermore, the combination of resonant absorption (and phase mixing) and the KHI leads to anti-phase behaviour between the line-of-sight (LOS) velocity and the (transverse) plane-of-the-sky (POS) motion, a characteristic observed recently in a prominence by \textit{Hinode}/SOT and \textit{IRIS} \citep{Okamoto_2015ApJ...809...71O}. 

Here we show that the combination of resonant absorption and the KHI can be readily seen in current coronal imaging instruments as small amplitude decay-less transverse oscillations, thus providing an explanation for the recent observations. We further demonstrate a potential MHD seismology application. 

\section{Numerical Model}\label{model}

Our 3D MHD numerical model is the same as in \citet{Antolin_2014ApJ...787L..22A}, where we take a loop with a density and temperature contrast with respect to the ambient low-$\beta$ coronal atmosphere. The loop is initially in hydrostatic equilibrium and has density and temperature ratios $\rho_i/\rho_e=3$ and $T_i/T_e=1/3$, respectively, where the index $i$ ($e$) denotes internal (external) values. The magnetic field is uniformly set to $B=22.8$~G. We take initially $T_i=10^6~$K, and $\rho_i=3\times10^9 \mu m_{p}$~g~cm$^{-3}$ ($\mu=0.5$ and $m_p$ is the proton mass). The loop boundary has a width of $\ell/R\approx0.4$, where $R$ is the radius of the loop. The length $L$ of the loop is $200~R$, and we set $R=1~$Mm.

The loop is subject to a perturbation mimicking a fundamental kink mode (longitudinal wavenumber $kR=\pi R/L\approx$0.015) with an initial amplitude of $v_0 =0.05~c_s$, with $c_s$ the external sound speed. This corresponds to $v_0=$15~km~s$^{-1}$ in our model. The corresponding kink phase speed is $c_k\approx1574~$km~s$^{-1}$. For further details please refer to \citet{Antolin_2014ApJ...787L..22A}.

We perform the 3D MHD simulation described above with the CIP-MOCCT scheme \citep{Kudoh_1999_CFD.8} including constant resistivity and viscosity. The MHD equations are solved, excluding gravity and loop curvature, which are second order factors for the present work. Furthermore, the effects of radiative cooling and thermal conduction are also neglected, as they are expected to be unimportant due to their longer timescales compared to that of the kink wave period. 

The numerical box is 512 $\times$ 256 $\times$ 50 points in the $x, y$ and $z$ directions respectively, where $(x,y)$ denote the transverse plane to the loop axis (along $z$) and $x$ is the direction of oscillation. Thanks to the symmetric properties of the kink mode only half the plane in $y$ and half the length of the loop are modelled (from $z=0$ to $z=100~$R), and we set symmetric boundary conditions in all boundary planes except for the $x$ boundary planes, where periodic boundary conditions are imposed. In order to minimise the influence from side boundary conditions (along $x$ and $y$), the spatial grids in $x$ and $y$ are non-uniform, with exponentially increasing values for distances well beyond the maximum displacement. The maximum distance in $x$ and $y$ from the centre is $\approx 16~R$. The spatial resolution at the loop's location is 0.0156~R$=15.6~$km. From a parameter study, we estimate that the effective (combined explicit and numerical) Reynolds and Lundquist numbers in the code are of the order of $10^4-10^5$.

\section{Numerical results and forward modelling}\label{results}

\subsection{A dual mechanism leading to strand-like structure in EUV emission lines}

Following the initial perturbation, the loop starts oscillating with a period $P=256~$s, closely corresponding to the period of the fundamental mode $2L/c{_k}$. The maximum displacement of the loop is 440~km ($0.44~R$), leading to a transverse velocity of $6.9~$km s$^{-1}$. 

The kink mode produces a velocity shear at the boundary of the flux tube, particularly between the amplified (azimuthal) resonant flow and the purely transverse motion of the kink mode, which generates the KHI. The KHI enhances the mixing with the external plasma and the viscous and magnetic dissipation of the kinetic and magnetic energy from the narrow resonant region. TWIKH rolls occur all along the flux tube in a roughly synchronised manner and are able to compress the plasma. They therefore significantly deform the density and temperature structure in the transverse plane of the flux tube. Such vortices are therefore regions of enhanced emissivity resulting in clear strand-like structure in EUV intensity images of the loop, as first shown in \citet{Antolin_2014ApJ...787L..22A} \citep[and thread-like structure in chromospheric conditions,][]{Antolin_2015ApJ...809...72A}.

\subsection{Emission lines, line-of-sight (LOS) and resolution}\label{forward}

For comparison with observations we forward model the results from the numerical simulation into observable quantities using the FoMo code \citep{VanDoorsselaere_0.3389/fspas.2016.00004}. We take the LOS plane perpendicular to the loop axis and define the LOS angle such that $0^{\circ}$ is perpendicular to the axis of oscillation. We choose the \ion{Fe}{9}~171~\AA~and \ion{Fe}{12}~193~\AA~emission lines, which have maximum formation temperatures of $\log T=5.9$ and $\log T=6.2$, respectively. Accordingly, in this particular model, the \ion{Fe}{9}~171~\AA~line is more tuned to detect the plasma response at the core of the loop, while the \ion{Fe}{12}~193~\AA~line is more sensitive to the hotter plasma near the boundaries (right hand panels of Fig.~\ref{fig1}). From here on, we refer to the \ion{Fe}{9}~171~\AA~line as the core line, and the \ion{Fe}{12}~193~\AA~line as the boundary line. 

For correct comparison with observations with a given instrument with resolving power of X (defined as the FWHM of the PSF and we take the plate-scale equal to half the spatial resolution) we degrade the original spatial resolution of the numerical model by first convolving the image of interest with a Gaussian with FWHM of X. We then resample the data according to the specific pixel size of the target instrument and add photon noise (Poisson distributed).  

\subsection{EUV variability}

\begin{figure}[!ht]
\begin{center}
$\begin{array}{c}
\includegraphics[scale=0.8]{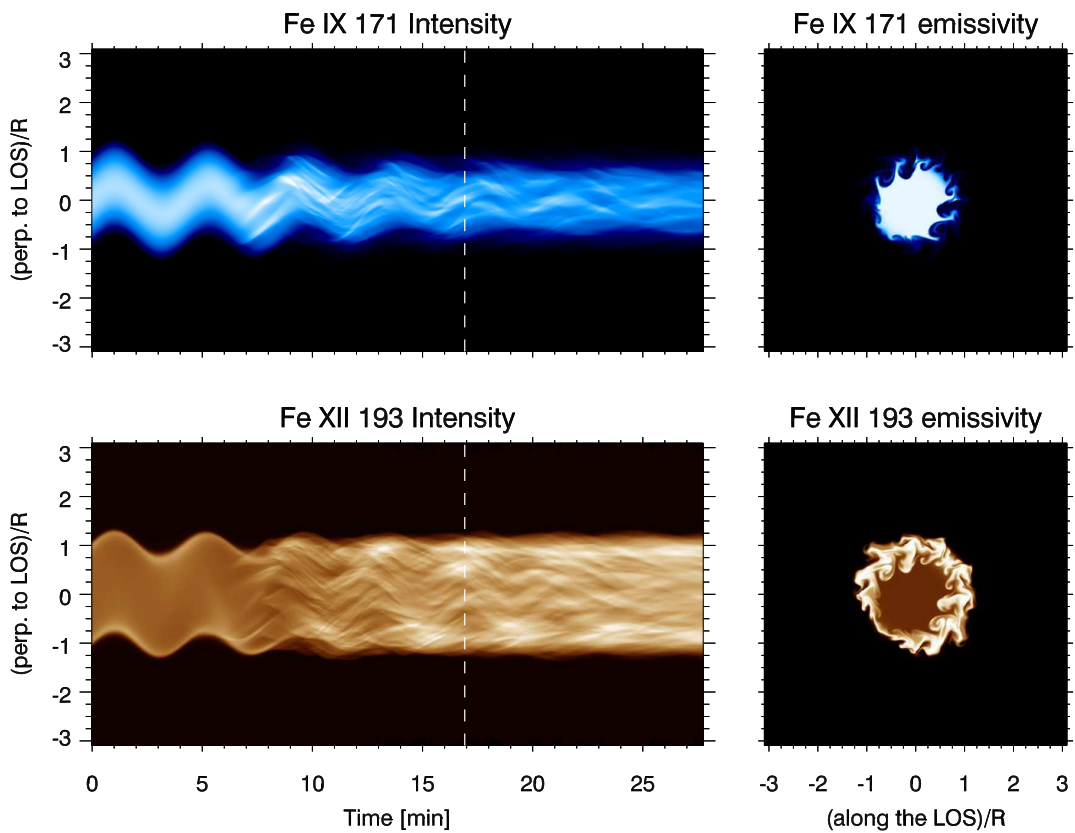}
\end{array}$
\caption{\textit{Left panels}: Time-distance diagrams of the forward modelling of the numerical model in the \ion{Fe}{9}~171~\AA~(in blue, top panel) and \ion{Fe}{12}~193~\AA~(in brown, bottom panel) intensity for a slit placed perpendicular at the apex of the loop, at a $45^{\circ}$ LOS angle and at numerical (highest) spatial resolution.  \textit{Right panels:} Snapshot of the cross-section of the emissivity for each line ($G_\lambda(T,n_e)n_e^2$, with $G$ the contribution function and $n_e$ the electron density) for the time indicated by the dashed line in the time-distance diagrams. The cross-section is rotated by the same LOS angle. 
\label{fig1}}
\end{center}
\end{figure}

Bright oscillatory strand-like structures are noticeable in Fig.~\ref{fig1} in both emission lines after about 2 periods, corresponding to the formation of TWIKH rolls. The amount of strand-like structure is largest in the boundary line. The overall damping of the kink mode is clear in \ion{Fe}{9}~171~\AA~. However, in \ion{Fe}{12}~193~\AA, it becomes hard to discern, partly because of the large amount of fine-scale structure produced by the TWIKH rolls. A careful look into this substructure reveals inner oscillations damping to a much weaker extent.

While the loop is observed to become thinner ($\approx15\%$) and dimmer ($\approx20\%$) in time in \ion{Fe}{9}~171~\AA, in \ion{Fe}{12}~193~\AA~the loop becomes broader ($\approx25\%$) and significantly brighter ($\approx100\%$). The brightness increase and broadening occurs rapidly, in about one period after the onset of the KHI. Periodic brightening is also observed, mostly in the boundary line close to the edges of the loop. These localised enhancements, caused by LOS superposition of the TWIKH rolls, perturb the overall intensity trend by about 1.5\% and 0.2\% for \ion{Fe}{12} and \ion{Fe}{9}, respectively. 

\subsection{Imaging characteristics at low spatial resolution and different LOS}\label{decayless}

\begin{figure}[!ht]
\begin{center}
$\begin{array}{c}
\includegraphics[scale=0.5]{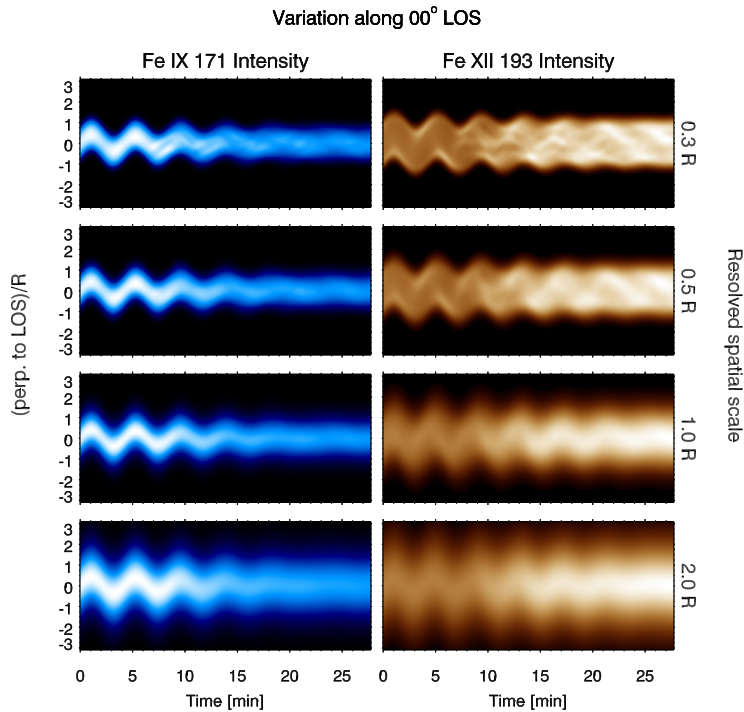}
\end{array}$
\caption{Time-distance diagram of the \ion{Fe}{9} (\textit{left column}) and \ion{Fe}{12} (\textit{right column}) intensities for the same configuration as in Fig.~\ref{fig1} but for a $0^{\circ}$ LOS and different spatial resolution (indicated on the right-hand side axis of the right column), obtained through Gaussian convolution only (see section~\ref{forward} for details). 
\label{fig2}}
\end{center}
\end{figure}

\begin{figure}[!ht]
\begin{center}
$\begin{array}{c}
\includegraphics[scale=0.6]{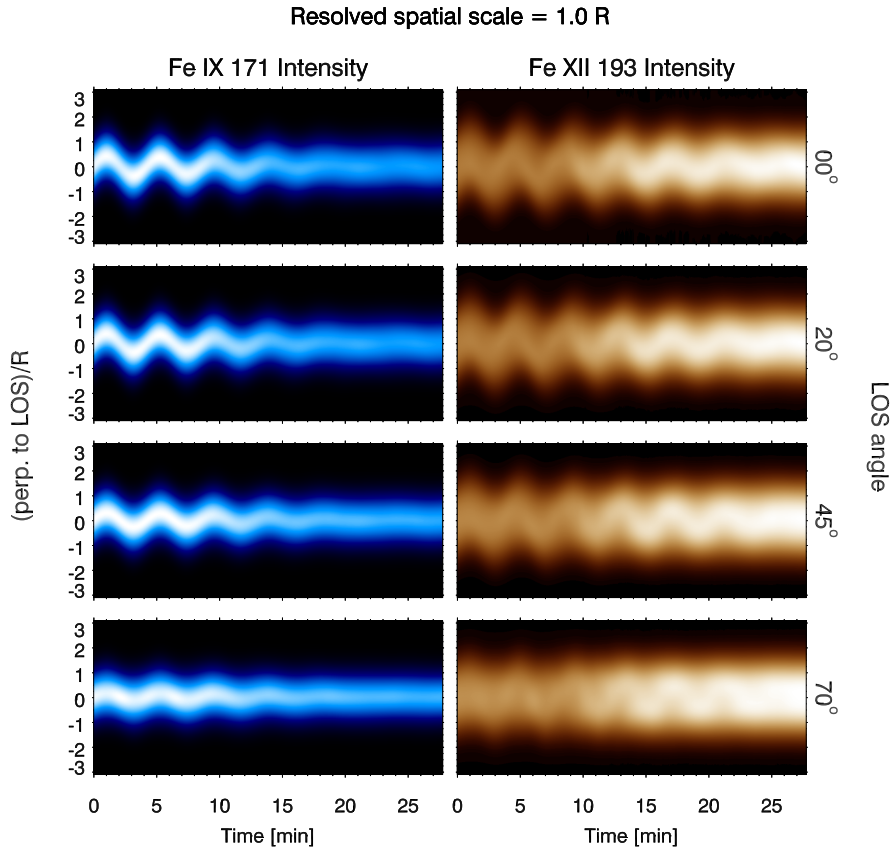}
\end{array}$
\caption{Similar to Fig.~\ref{fig2} but for a fixed spatial resolution of $1~R$ and various LOS angles (indicated on the right-hand side axis of the right column).
\label{fig3}}
\end{center}
\end{figure}

As the spatial resolution is degraded, Fig.~\ref{fig2} shows that the fine-scale strand-like structure completely vanishes for a resolved spatial scale of $0.5~R$, although some features can still be noticed in the boundary line. For lower spatial resolutions, only the intensity variation and overall damping behaviour of the kink mode is observed. Contrary to the core line, in the boundary line, besides the brightening, the damping appears negligible. The apparent decay-less oscillation can appear non-sinusoidal, which is particularly clear in the third right panel from top in Fig.~\ref{fig2} in the first 4 oscillations. Figure~\ref{fig3} shows that these effects are independent of the LOS angle. 

\begin{figure}[!ht]
\begin{center}
$\begin{array}{c}
\includegraphics[scale=0.4]{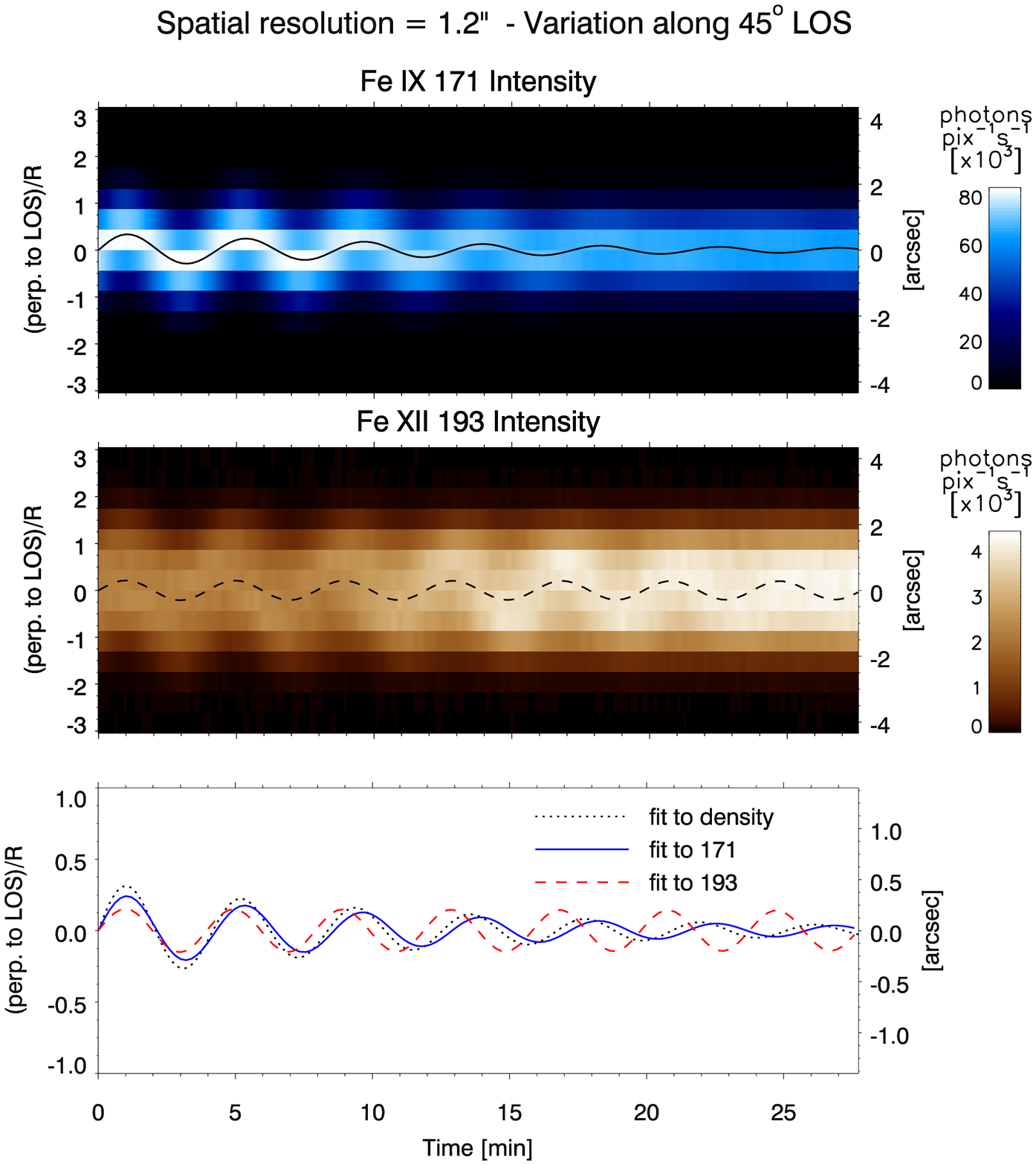}
\end{array}$
\caption{Time-distance diagrams for a slit in the same configuration as in Fig.~\ref{fig1}, targeting an imaging instrument such as \textit{SDO}/AIA. The damping profile is overlaid on each diagram, calculated by first fitting a Gaussian to the profile at each time step from which the centroid is obtained. Then we fit an exponentially damped cosine to the centroid locations rather than a Gaussian damping profile since the transition from Gaussian to exponential damping is expected after 2 periods \citep[][also, this allows direct comparison to the observational reports]{Pascoe_2015AA...578A..99P}. The fits are reproduced in the lower panel, in which the damping to the density is also shown. The latter is produced by integrating the density cross-section at the loop apex along the same LOS ray (45$^\circ$), removing the background component and by degrading the spatial resolution to $1.2\arcsec$. We then apply the same fitting procedure as for emission lines. 
\label{fig4}}
\end{center}
\end{figure}

To further improve comparison with \textit{SDO}/AIA observations, we convolve and rebin the results to achieve a spatial resolution of $1.2\arcsec$ (as explained in section~\ref{forward}). As can be seen in Fig.~\ref{fig4}, the overall intensity variation, the damping in the core line and decay-less oscillation in the boundary line can still be clearly observed. The observed displacement of the loop in the core and boundary lines is $370$~km and $230$~km, respectively.
 
In the bottom panel of Fig.~\ref{fig4} we notice that the boundary line profile starts oscillating in-phase with the core line and the density but goes quickly out-of-phase, an effect clearly noticeable after 2 periods. In particular, the oscillatory profile of the core line goes above $90^\circ$ out-of-phase with the boundary line after about 4 periods. The fits to the core and boundary lines give periods of 258~s and 240~s, respectively. The fit to the density gives a period of 253~s. 

\section{Discussion and Conclusions}\label{discussion}

In this paper, we have investigated imaging signatures of the transverse MHD wave derived from 3D MHD numerical simulations and forward modelling. The main factors affecting the intensity modulation in the host loops are the KHI and resonant absorption. Importantly, the KHI vortices (TWIKH rolls) carry over characteristics of the resonant absorption and phase-mixing mechanisms, allowing these to be detected with current instrumentation. We have found that the KHI produces significant intensity changes and that emission lines with different temperature sensitivity to either the loop core or the external medium provide insights into different wave modes.

At high spatial resolution and especially in the hot line, the fine strand-like structure generated by the TWIKH rolls can be observed, as described in \citet{Antolin_2014ApJ...787L..22A}. At a lower spatial resolution corresponding to AIA, we have shown that the detected damping and period can vary depending on the emission line, which can lead to an out-of-phase behaviour between core and boundary lines. These results can be understood from the fact that the boundary line is more sensitive to the KHI dynamics, to resonant absorption and phase mixing. A decay-less oscillation is obtained in the boundary line for any LOS angle and this is the result of periodic brightenings and the coherent motion of the vortices. The vortices, in turn, result from unstable azimuthal Alfv\'en waves. Therefore, this decay-less oscillation reflects more the Alfv\'en waves coupled resonantly to the kink mode than the global kink mode itself. Resonant absorption transfers the energy of the kink mode to the Alfv\'en waves, whose damping is expected to be much lower than that of the kink mode since it relies on phase mixing and the turbulence resulting from the instability. The Alfv\'en waves act therefore as energy reservoir for the TWIKH rolls, which persist over time leading to the decay-less oscillation.

A shorter period is found for the boundary line and is due to phase mixing. Indeed, the azimuthal Alfv\'en waves in the boundary layer have an increasingly higher phase speed the farther they are from the loop core. Since it is these waves that are enhanced due to resonant absorption and which become K-H unstable, their signal becomes dominant in the boundary line. The fit to the density from the numerical model ends up with a shorter period than that obtained from the core line. This is because the density profile is influenced by the KHI dynamics, which broaden the flux tube, even though the small-scale structure produced by the instability is unresolved. Accordingly, when fitting only the central section of the flux tube along the oscillation axis, which is minimally influenced by resonant absorption and by the KHI, we obtain a period of 256~sec, closer to that obtained from the core line. 

The out-of-phase behaviour between the core and boundary lines is ultimately the same effect as that leading to the out-of-phase behaviour between the LOS velocity and the POS motion seen in the core line \citep{Antolin_2015ApJ...809...72A}, explaining the observations by \citet{Okamoto_2015ApJ...809...71O}. The difference in oscillation period between core and boundary lines can be used to obtain an estimate of the density at location of emission in the boundary layer and also to estimate the density contrast between the loop and the external corona. The period observed with the boundary line (here \ion{Fe}{12}~193~\AA) satisfies $P_{b}=2L/v_{A_b}$, corresponding to the period of azimuthal Alfv\'en waves dominating the signal, propagating with speed $v_{A_b}=B/\sqrt{4\pi\rho_b}$. Then $\rho_{b}=B^2 P_{b}^2/(16\pi L^2)$. Since the plasma $\beta$ is low, we can approximate the kink speed as $c_k=\sqrt{2/(1+\rho_e/\rho_i)}v_{A_i}$ and we have $P_k=2L/c_k$, where $P_k$ is the period of the kink speed, which is to a large accuracy the observed period with the core line. Using these equations to replace the magnetic field we obtain:
\begin{equation}\label{rob2}
\frac{\rho_b}{\rho_i}=\frac{1}{2}\left(1+\frac{\rho_e}{\rho_i}\right)\left(\frac{P_b}{P_k}\right)^2.
\end{equation}
Replacing with the observed values for $P_b$ and $P_k$, and assuming that $\rho_e$ and $\rho_i$ are known, we obtain $\rho_b\approx1.73\times10^{9}\mu m_p$~g~cm$^{-3}$. The average location (in time and angles) in the boundary corresponding to this density has a temperature of $1.83\times10^{6}~$K, very close to the maximum formation temperature of the line. From Eq.~\ref{rob2} we can provide an upper limit to the density contrast:
\begin{equation}\label{rob3}
\frac{\rho_e}{\rho_i}<\min\left\{\frac{(P_b/P_k)^2}{2-(P_b/P_k)^2}, \frac{2-(P_b/P_k)^2}{(P_b/P_k)^2}\right\}
\end{equation}
which gives a lowest upper limit of 0.76 (the true value being 0.34).

These results open several possibilities for MHD seismology of the loop shell structure. Indeed, the KHI  broadens the density profile, regardless of its initial shape, and the plasma emissivity at a particular temperature in the boundary ends up coming from a small dynamic range. Multiple periods can thus be detected using multiple channels sensitive to different temperatures in the boundary, leading to a temperature dependent density and the radial structure of a loop \citep[as suggested also by][]{Verth_2010ApJ...714.1637V, Fedun_2011ApJ...740L..46F}. Therefore, this technique also allows to probe the difference in temperature between loops and their surroundings.

Our model assumes that the loop is colder than the ambient corona. Note that this model differs minimally from a model in which the loop is both hotter and denser than the ambient corona. Indeed, both the resonant properties and KHI onset in the loop remain largely the same. For example, in a model with an ambient temperature of $1$~MK and an internal loop temperature of $1.5$~MK, similar forward modelling results would be obtained with the same pair of emission lines, except that the results corresponding here to the core and boundary lines would be switched. The loop core and boundary would be better visualised with the \ion{Fe}{12} line and the \ion{Fe}{9} line, respectively.

The observed transverse displacement and periods in our model match those observed exhibiting the decay-less kink mode oscillations in coronal loops \citep{Nistico_2013AA...552A..57N,Anfinogentov_2013AA...560A.107A,Anfinogentov_2015AA...583A.136A,Goddard_2016AA...585A.137G}. Our results therefore provide an explanation for these observations. Unfortunately in these studies only the \ion{Fe}{9}~171~\AA~line has been used and it is uncertain how multi-thermal these loops are and what the temperature difference is with the external corona. The loops shown in these studies present significant intensity changes during their oscillation, in a similar way to the changes observed in our model. Furthermore, the oscillations in our model can appear non-sinusoidal, especially in the first few periods. Such non sinusoidal and rather pointy oscillations can be seen in several of the examples provided by \citet{Anfinogentov_2013AA...560A.107A,Anfinogentov_2015AA...583A.136A,Goddard_2016AA...585A.137G}.

The presence of the KHI and resonant absorption have been demonstrated to be quite robust with respect to density and longitudinal magnetic field variation across the loop, perturbation amplitudes, thickness of the boundary layer, longitudinal flows and magnetic twist \citep{Antolin_2014ApJ...787L..22A, Murawski_2016MNRAS.459.2566M, Terradas_etal_2016}. The described decay-less effect is however limited and will not be observed for strong perturbation amplitudes. This is because the TWIKH rolls that are associated with resonant absorption are always around the flux tube. The combined effect from the coherent motion of vortices and periodic brightening is therefore confined to a transverse layer equal to the loop width, which may explain why such decay-less oscillations are only seen for low amplitudes. On the other hand, for large amplitudes a negative correlation between damping time and perturbation amplitude has been found by \citet{Goddard_2016AA...590L...5G}, suggesting that non-linear effects such as those described here may still play an important role in the observed overall dynamics of the loop.

\acknowledgments
This research has received funding from the UK Science and Technology Facilities Council and the European Union Horizon 2020 research and innovation programme (grant agreement No. 647214), and also from JSPS KAKENHI Grant Numbers 25220703 (PI: S. Tsuneta) and 15H03640 (PI: T. Yokoyama). T.V.D. was supported by FWO Vlaanderen's Odysseus programme, GOA-2015-014 (KU Leuven) and the IAP P7/08 CHARM (Belspo). Numerical computations were carried out on Cray XC30 at the Center for Computational Astrophysics, NAOJ. This work also benefited from the ISSI - Coronal Rain (P.I. Patrick Antolin) and ISSI-BJ meetings (P.I. V. Nakariakov \& T. Van Doorsselaere).

\bibliographystyle{aasjournal}
\bibliography{ms.bbl}

\end{document}